\documentclass[prd,aps,floats,epsfig,twocolumn,showpacs]{revtex4}  

\usepackage{enumitem}
\usepackage{graphicx}
\usepackage{subfigure}
\usepackage{amssymb}
\usepackage{amsmath}
\usepackage{bbm}
\usepackage{color}

\input{epsf}

\def\lsim{\mathrel{\mathop
  {\hbox{\lower0.5ex\hbox{$\sim$}\kern-0.8em\lower-0.7ex\hbox{$<$}}}}}
\def\gsim{\mathrel{\mathop
  {\hbox{\lower0.5ex\hbox{$\sim$}\kern-0.8em\lower-0.7ex\hbox{$>$}}}}}

\def\lapprox{\hbox{\lower .8ex\hbox{$\,\buildrel < \over\sim\,$}}}
\def\gapprox{\hbox{\lower .8ex\hbox{$\,\buildrel > \over\sim\,$}}}

\newcommand{\be}{\begin{equation}}
\newcommand{\ee}{\end{equation}}
\newcommand{\ba}{\begin{aligned}}
\newcommand{\ea}{\end{aligned}}
\newcommand{\bea}{\begin{eqnarray}}
\newcommand{\eea}{\end{eqnarray}}

\begin{document}
\title{Magnification Bias Corrections to Galaxy-Lensing Cross-Correlations}

\author{Riad Ziour$^{1}$ and Lam Hui$^{2,3}$}
\affiliation{
$^{1}$Laboratoire APC, UMR7164 (Univ.~Paris7, CNRS, CEA, Obs. deParis),10 rue A. Domon et L. Duquet, 75205 Paris Cedex 13, France. \\
$^{2}$Institute for Strings, Cosmology and Astroparticle Physics (ISCAP)\\
$^{3}$Department of Physics, Columbia University, New York, NY 10027, U.S.A.\\
ziour@apc.univ-paris7.fr, lhui@astro.columbia.edu
}

\date{\today}

\begin{abstract}
Galaxy-galaxy or galaxy-quasar lensing can provide important information on the mass distribution in the universe. 
It consists of correlating the lensing signal (either shear or magnification) of a background galaxy/quasar sample 
with the number density of a foreground galaxy sample.
However, the foreground galaxy density is inevitably altered by the magnification bias due to the mass 
between the foreground and the observer, leading to a correction to the observed galaxy-lensing signal. 
The aim of this paper is to quantify this correction. 
The single most important determining factor is
the foreground redshift $z_f$: the correction is small if the foreground galaxies are at
low redshifts but can become non-negligible for sufficiently high redshifts. 
For instance, we find that for the multipole $\ell=1000$, the correction is above $1\%\times (5s_{f}-2)/b_{f}$ for $z_{f}\gtrsim 0.37$, and above $5\%\times (5s_{f}-2)/b_{f}$ for $z_{f}\gtrsim 0.67$, where $s_{f}$ is the number count slope of the foreground sample, 
and $b_{f}$ its galaxy bias. 
These considerations are particularly important for geometrical measures, such as the Jain and Taylor
ratio or its generalization by Zhang et al. 
Assuming $(5s_f - 2)/b_f = 1$, 
we find that the foreground redshift should be limited to $z_f \lesssim 0.45$ in order to
avoid biasing the inferred dark energy equation of state $w$ by more
than $5\%$, and that even for a low foreground redshift ($< 0.45$), the background samples must
be well separated from the foreground to avoid incurring a bias of similar magnitude.
Lastly, we briefly comment on the possibility of obtaining these geometrical measures without
using galaxy shapes, using instead magnification bias itself.
\end{abstract}

\pacs{98.80.-k; 98.80.Es; 98.65.Dx; 95.35.+d}


\maketitle
\section{Introduction}

The gravitational lensing of distant objects such as galaxies or quasars \footnote{In this article, we will use the 
term \emph{galaxy} as a generic term for any survey object like actual galaxies or quasars.} located around a redshift $z_{b}$ 
consists of both shear (a change in shape) and magnification (a change in size and flux), and is due to the integrated mass along the 
line of sight (LOS) between these background galaxies and the observer. 
Correlating this lensing signal (either shear or magnification) with the foreground galaxy number density at $z_{f}<z_{b}$, 
we pick up the only  part of the lensing mass which is correlated with the galaxy density at $z_{f}$, \textit{i.e.} the matter 
located around $z_{f}$. Galaxy-lensing correlations are therefore very interesting tools to study the matter distribution 
around the lenses.
These correlations have been observed experimentally and studied theoretically both in the case of galaxy-shear correlation 
(also often called galaxy-galaxy correlations) \cite{Sheldon:2004,Sheldon:2007_I,Johnston:2007_II,Sheldon:2007_III,Guzik_Seljak:2001,Guzik_Seljak:2002,Mandelbaum:2005,Mandelbaum:2006} and galaxy-magnification correlation \cite{Gunn:1967,Turner:1984,Webster1988,Fugmann:1988,Narayan:1989,Schneider:1989,Broadhurst:1994,Villumsen:1995,Villumsen:1997,Moessner_JV:1998,Moessner_J:1998,Gaztanaga:2002,Scranton_SDSS:2005,Menard:2002,Jain_S:2003,Gaztanaga2008}. In addition to providing information on the matter distribution, galaxy-lensing 
cross-correlations can be combined in order to build quantities which depend only on geometrical distances to the 
background and foreground samples, allowing measurements of dark energy (DE) parameters independently of the exact 
expression of the matter power spectrum. These methods, that we call \emph{geometrical methods}, have been 
presented in \cite{Jain_T:2004,Taylor:2006,Bernstein:2003} in the case of galaxy-shear correlations and in a 
more general way in \cite{Zhang_Hui:2003} for both galaxy-shear and shear-shear correlations. 

In this article, we would like to draw attention to the fact that these galaxy-lensing correlations depend crucially
on estimates of the foreground galaxy density, which is inevitably altered by magnification bias, leading
potentially to a systematic source of error. Magnification bias
is a well known effect (a derivation can be found in \cite{Hui_Anis_I:2007}). 
Corrections due to magnification bias caused by the presence of the mass 
along the LOS between $z_{f}$ and $z={0}$ enter as follows. Suppose there is a mass overdensity along the LOS. 
On the one hand, the observed number of galaxies per unit 
area on the sky is smaller due to the stretching of the apparent inter-galaxy spacing; on the other hand, very faint 
galaxies which were under the threshold of detectability are now visible, leading to a higher number density. 
The net effect depends on the number count slope $s_{f}$ and lead to a correction term being added 
to the intrinsic galaxy density
\be
\delta_{n}=\delta_{g}+\delta_{\mu}
\ee
where $\delta_{\mu}\propto (5s_{f}-2)$. As a consequence, the galaxy-lensing cross-correlation gets an extra term due 
to the correlation between the lensing signal of the background galaxies and the extra foreground galaxy density 
$\delta_{\mu}$. How important this effect is, and under what condition this should be taken into account, is
addressed as follows.

In section \ref{Formalism and correlation functions}, we present the expressions for the magnification bias 
and for the galaxy-lensing correlations, including both galaxy-shear and galaxy-magnification correlations \footnote{A clarification on possibly confusing terminology: we use the term galaxy-lensing correlations
to refer to both galaxy-shear and galaxy-magnification correlations. Measurements of galaxy-shear correlations
using a background galaxy sample (for shear) and a foreground galaxy sample (for counts) 
are often referred to as galaxy-galaxy lensing
in the literature.
Measurements of galaxy-magnification correlations using a background quasar sample (for counts) 
and a foreground galaxy sample (for counts) are sometimes referred to as galaxy-quasar associations.
}.
Then, 
we study in section \ref{Correction due to magnification bias} how the correction due to magnification bias depends 
on various parameters in the problem: the background redshift $z_b$, the foreground redshift $z_f$,
the angular scale, the selection function width $\sigma$, the number count slope $s_f$ and the galaxy bias $b_f$.
In section \ref{Geometrical methods}, we point out that magnification bias corrections are
particularly important for the geometrical methods of \cite{Jain_T:2004} 
and \cite{Zhang_Hui:2003}. We discuss strategies for minimizing them. We also
suggest a simple extension of these geometrical methods which uses magnification bias itself,
without the need for any galaxy shape information.
We conclude in section \ref{Conclusion}.
One recurring theme that would come up in our discussion is that the foreground
redshift $z_f$ is an important determining factor in the size of the magnification bias
corrections. This is not only because a larger $z_f$ makes for a larger lensing efficiency,
it is also because a larger $z_f$ generally means intrinsically brighter galaxies and therefore
steeper number count slope $s_f$. The latter is particulary true if precise (i.e. spectroscopic) redshifts
are necessary for the foreground galaxies, such as in some versions of the geometrical methods.

\section{Formalism and correlation functions}\label{Formalism and correlation functions}
\subsection{Magnification bias}
When magnification bias is taken into account, the observed galaxy overdensity at a mean redshift $z_{0}$ in the direction $\boldsymbol{\theta}$ is 
\be\label{delta n}
\delta_{n}(\boldsymbol{\theta},z_{0})=\delta_{g}(\boldsymbol{\theta},z_{0})+\delta_{\mu}(\boldsymbol{\theta},z_{0})\ .
\ee
The first term is the intrinsic galaxy fluctuation integrated over a normalized window function $W(z,z_{0})$ and is given by
\be
\delta_{g}(\boldsymbol{\theta},z_{0})=\int_{0}^{\infty}dz\; W(z,z_{0})b(z)\delta[\chi(z)\boldsymbol{\theta},z]
\ee
where $\chi(z)$ is the comoving distance to redshift $z$ and $\delta=(\rho-\bar{\rho})/\bar{\rho}$ is the matter overdensity. We consider here the case of a flat universe, but the expression can be easily generalized to an open or closed universe. We assume a scale independent bias such that $\delta_{g}=b(z)\delta$; this should be correct for $k\lesssim 0.05 h \text{Mpc}^{-1}$, though corrections could become important at smaller scale (see \textit{e.g.} \cite{Smith_Scoccimarro:2006}). If 
the bias $b$ varies slowly across the width of the selection function $W$, we can further assume that the bias is constant through the galaxy sample, allowing the simplification
\be
\delta_{g}(\boldsymbol{\theta},z_{0})=b(z_{0})\int_{0}^{\infty}dz\; W(z,z_{0})\delta[\chi(z)\boldsymbol{\theta},z] \ ;
\ee
we will assume the relation holds in this paper, although it is straightforward to generalize our results to
encompass both scale dependent and rapidly evolving bias.

The second term of Eq.(\ref{delta n}) encodes the overdensity due to magnification bias, and reads (assuming that $s(z)$ is slowly varying across $W(z,z_{0})$):
\be\label{delta mu}
\delta_{\mu}(\boldsymbol{\theta},z_{0})=(5s(z_{0})-2)\int_{0}^{\infty}\!\!dz \frac{c}{H(z)} g(z,z_{0})\nabla_{\bot}^2\phi\left[\chi(z)\boldsymbol{\theta},z\right]
\ee
where $\nabla_{\bot}^2\phi$ is the 2D Laplacian of the gravitational potential in the plane perpendicular to the line of sight and $s$ is the slope of the number count function (see \textit{e.g.} \cite{Hui_Anis_I:2007}). For a survey with limiting magnitude $m$ this is
\be
s=\frac{d\log_{10}N(<m)}{dm}\ .
\ee
We have also introduced the lensing weight function
\be\label{function g}
 g(z,z_{0})=\chi(z)\int_{z}^{\infty}dz' \frac{\chi(z')-\chi(z)}{\chi(z')}W(z',z_{0})
\ee
that can be thought of as proportional to the probability for a source around $z_{0}$ to be lensed by mass at redshift $z$; it is peaked at a redshift corresponding roughly to half the comoving distance to the source (for a review of weak lensing, see \textit{e.g.} \cite{Bartelmann:1999}).

If we make use of the Poisson's equation, Eq.(\ref{delta mu}) becomes
\be\label{delta mu final}
\begin{aligned}
\delta_{\mu}(\boldsymbol{\theta},z_{0})=\left(\frac{3}{2}\frac{H_{0}^{2}}{c^{2}}\Omega_{m,0}\right)(5s(z_{0})-2)\qquad\quad\quad\quad\\
\times\int_{0}^{\infty}dz \frac{c}{H(z)} g(z,z_{0})(1+z)\delta\left[\chi(z)\boldsymbol{\theta},z\right].
\end{aligned}
\ee

\subsection{Cross-correlations}
\subsubsection{Galaxy-shear correlations}
\label{gshearcorr}
Galaxy-shear correlation (also often called galaxy-galaxy lensing) has been the object of considerable interest from theoretical and observational points of view \cite{Sheldon:2004,Sheldon:2007_I,Johnston:2007_II,Sheldon:2007_III,Guzik_Seljak:2001,Guzik_Seljak:2002,Mandelbaum:2005,Mandelbaum:2006}. It consists of correlating the cosmic shear of background galaxies located around a redshift $z_{b}$ with the number density of foreground galaxies at redshift $z_{f}$. Here, we focus on the electric-component of the shear (sometimes called the scalar component or the gradient mode) and we will use $\kappa$ to refer to this scalar part: the convergence. The cosmic shear of the sources is due to all the mass between $z_{b}$ and $z=0$; however correlating this shear with galaxy number density at $z_{f}$, we extract the part of the lensing mass that is correlated with the galaxy density at $z_{f}$, \textit{i.e.} the matter located around $z_{f}$. Therefore, galaxy-shear correlations allow us to get important information about the galaxy-mass power spectrum.

More precisely, due to magnification bias, the galaxy-shear two-point correlation function is the sum of two terms
\be
\begin{aligned}\label{w n gamma}
w_{n\kappa }(\theta, z_{f},z_{b})&\equiv\left<\delta_{n}(\boldsymbol{\theta},z_{f})\kappa(\boldsymbol{\theta}',z_{b})\right> \\
&=\left<\delta_{g}(\boldsymbol{\theta},z_{f})\kappa(\boldsymbol{\theta}',z_{b})\right>+\left<\delta_{\mu}(\boldsymbol{\theta},z_{f})\kappa(\boldsymbol{\theta}',z_{b})\right>\\
&=w_{g\kappa }(\theta, z_{f},z_{b})+w_{\mu\kappa }(\theta, z_{f},z_{b})\qquad
\end{aligned}
\ee
where $\cos \theta=\boldsymbol{\theta}.\boldsymbol{\theta}'$. We then consider the Legendre coefficients $C^{g\kappa}_{\ell}(z_{f},z_{b}),C^{\mu\kappa}_{\ell}(z_{f},z_{b}) $ of the correlation functions. They are defined as
\be
w_{g\kappa }(\theta, z_{f},z_{b})=\sum_{l}\frac{2\ell+1}{4\pi}C^{g\kappa}_{\ell}(z_{f},z_{b})P_{\ell}(\cos \theta)
\ee
where the $P_{\ell}$ are the Legendre polynomials. The galaxy-shear angular power spectrum will then also be the sum of two terms,
\be
C^{n\kappa}_{\ell}(z_{f},z_{b})=C^{g\kappa}_{\ell}(z_{f},z_{b})+C^{\mu\kappa}_{\ell}(z_{f},z_{b}) .
\ee
We calculate these two angular power spectra $C^{g\kappa}_{\ell},C^{\mu\kappa}_{\ell}$ using the Limber approximation which is accurate for large multipoles $\ell\gtrsim 10$ \cite{LoVerde:2008}:
\bea\label{Cggamma}
C^{g\kappa}_{\ell}(z_{f},z_{b}) =  \left(\frac{3}{2}\frac{H_{0}^{2}}{c^{2}}\Omega_{m,0}\right)b_{f}\nonumber\qquad\qquad\qquad\qquad\qquad\\
\times\int_{0}^{\infty}\!\!dz \frac{W(z,z_{f})g(z,z_{b})}{\chi^{2}(z)}(1+z)P(\ell/\chi(z),z)\nonumber\\
\\
C^{\mu\kappa}_{\ell}(z_{f},z_{b}) = \left(\frac{3}{2}\frac{H_{0}^{2}}{c^{2}}\Omega_{m,0}\right)^{2}(5s_{f}-2)\nonumber\qquad\qquad\qquad\\
\times\int_{0}^{\infty}\!\!dz  \frac{c}{H(z)}\frac{g(z,z_{f})g(z,z_{b})}{\chi^{2}(z)}(1+z)^{2}P(\ell/\chi(z),z)\label{Cmugamma}\nonumber\\
\eea
where $P(k,z)$ is the 3D matter power spectrum for a wave number $k$ and at a redshift $z$, $b_{f}$ the intrinsic clustering bias of the foreground galaxy and $s_{f}$ the mean number count slope of the foreground galaxy sample. Usually, the signal we are interested in is $C^{g\kappa}_{\ell}$, while $C^{\mu\kappa}_{\ell}$ will constitute the magnification bias induced correction to this signal. It is then important to know to what extent this correction can be neglected or should be taken into account. We will discuss these issues in detail in the following sections, but before doing so, we would like to emphasize that the same sort of corrections exist for galaxy-magnification cross-correlations.

\subsubsection{Galaxy-magnification correlations}\label{gmcorr}
Galaxy-magnification cross-correlation \cite{Gunn:1967,Turner:1984,Webster1988,Fugmann:1988,Narayan:1989,Schneider:1989,Broadhurst:1994,Villumsen:1995,Villumsen:1997,Moessner_JV:1998,Moessner_J:1998,Gaztanaga:2002,Scranton_SDSS:2005,Menard:2002,Jain_S:2003,Gaztanaga2008} is very similar to galaxy-shear correlation replacing the shear by the magnification of the background galaxies. In order to measure these correlations, one should correlate the galaxy densitiy of a background distribution with the density of a foreground distribution leading to the correlation function 
\be
\begin{aligned}
w_{nn}(\theta, z_{f},z_{b})=&w_{gg}(\theta, z_{f},z_{b})+w_{g\mu}(\theta, z_{f},z_{b})\\
&+w_{\mu g}(\theta, z_{f},z_{b})+w_{\mu\mu }(\theta, z_{f},z_{b}) \ .
\end{aligned}
\ee
If the background and foreground distributions do not overlap, and they are far apart, it is easy to see that the intrinsic galaxy-galaxy correlation $w_{gg}(\theta, z_{f},z_{b})$ and the correlation between magnification of foreground galaxies and background galaxy density $w_{\mu g}(\theta, z_{f},z_{b})$ both vanish. The two remaining terms 
\be
w_{nn}(\theta, z_{f},z_{b})=w_{g\mu}(\theta, z_{f},z_{b})+w_{\mu\mu }(\theta, z_{f},z_{b})
\ee
are comparable to Eq.(\ref{w n gamma}).

Similarly to $C^{n\kappa}_{\ell}$, the angular power spectrum will be the sum of two terms
\be
C^{n\mu}_{\ell}(z_{f},z_{b})=C^{g\mu}_{\ell}(z_{f},z_{b})+C^{\mu\mu}_{\ell}(z_{f},z_{b}) ,
\ee
where
\bea\label{Cgmu}
C^{g\mu}_{\ell}(z_{f},z_{b}) = \left(\frac{3}{2}\frac{H_{0}^{2}}{c^{2}}\Omega_{m,0}\right)b_{f}(5s_{b}-2)\qquad\qquad\qquad\nonumber\\
\times\int_{0}^{\infty}\!\!dz \frac{W(z,z_{f})g(z,z_{b})}{\chi^{2}(z)}(1+z)P(\ell/\chi(z),z)\nonumber\\
\\
C^{\mu\mu}_{\ell}(z_{f},z_{b}) = \left(\frac{3}{2}\frac{H_{0}^{2}}{c^{2}}\Omega_{m,0}\right)^{2}(5s_{f}-2)(5s_{b}-2)\qquad\nonumber\\
\times\int_{0}^{\infty}\!\!dz  \frac{c}{H(z)}\frac{g(z,z_{f})g(z,z_{b})}{\chi^{2}(z)}(1+z)^{2}P(\ell/\chi(z),z)\label{Cmumu}\nonumber\\
\eea
We can see that these expressions depend strongly on the galaxy samples, through the bias $b_{f}$ and the number count slopes $s_{f},s_{b}$.
Here, $C^{\mu\mu}_\ell$ constitutes the magnification bias correction to $C^{g\mu}_\ell$, the quantity that observers
hope to measure by cross-correlating number counts of two widely separated samples of galaxies.

\subsubsection{Formalism that combines the two types of correlations}\label{subsection Formalism that combines the two types of correlations}

If we compare Eq.(\ref{Cggamma}-\ref{Cmugamma}) and Eq.(\ref{Cgmu}-\ref{Cmumu}), it is easy to see that they have a similar structure. To give a unifying treatment, let's introduce the matter-convergence and convergence-convergence power spectra
\bea\label{Cgkappa}
C^{\delta\kappa}_{\ell}(z_{f},z_{b})=\left(\frac{3}{2}\frac{H_{0}^{2}}{c^{2}}\Omega_{m,0}\right)\qquad\qquad\qquad\qquad\qquad\nonumber\\
\times\int_{0}^{\infty}\!\!dz \frac{W(z,z_{f})g(z,z_{b})}{\chi^{2}(z)}(1+z)P(\ell/\chi(z),z),\nonumber\\
\\
C^{\kappa\kappa}_{\ell}(z_{f},z_{b}) = \left(\frac{3}{2}\frac{H_{0}^{2}}{c^{2}}\Omega_{m,0}\right)^{2}\qquad\qquad\qquad\qquad\qquad\nonumber\\
\times\int_{0}^{\infty}\!\!dz  \frac{c}{H(z)}\frac{g(z,z_{f})g(z,z_{b})}{\chi^{2}(z)}(1+z)^{2}P(\ell/\chi(z),z)\label{Ckappakappa};\nonumber\\
\eea
then, we can relate the different power spectra in the following way
\bea
C^{g\kappa}_{\ell}(z_{f},z_{b})&=& b_{f}\; C^{\delta\kappa}_{\ell}(z_{f},z_{b})\\
C^{\mu\kappa}_{\ell}(z_{f},z_{b})&=& (5s_{f}-2)\; C^{\kappa\kappa}_{\ell}(z_{f},z_{b})
\eea
and
\bea
C^{g\mu}_{\ell}(z_{f},z_{b})&=&(5s_{b}-2) b_{f}\; C^{\delta\kappa}_{\ell}(z_{f},z_{b})\\
C^{\mu\mu}_{\ell}(z_{f},z_{b})&=&(5s_{b}-2) (5s_{f}-2)\; C^{\kappa\kappa}_{\ell}(z_{f},z_{b}).
\eea
We are interested in the fractional magnification corrections to the observed galaxy-lensing correlations,
and in both the galaxy-shear correlation and the galaxy-magnification correlation, that ratio is
\bea
\label{def reff}
\frac{C^{\mu\kappa}_{\ell}(z_{f},z_{b})}{C^{g\kappa}_{\ell}(z_{f},z_{b})}
= \frac{C^{\mu\mu}_{\ell}(z_{f},z_{b})}{C^{g\mu}_{\ell}(z_{f},z_{b})}
= \frac{(5s_{f}-2)}{b_{f}}\; r(z_{f},z_{b};\ell)
\eea
where $r$ is defined as
\be\label{def r}
r(z_{f},z_{b};\ell)\equiv\frac{C^{\kappa\kappa}_{\ell}(z_{f},z_{b})}{C^{\delta\kappa}_{\ell}(z_{f},z_{b})}\; .
\ee
In what follows, we will display several figures showing $r$, but it should be kept in mind
that ultimately, the fractional correction of interest is $r$ multiplied by $(5s_f - 2)/b_f$,
a quantity which is sample dependent. It suffices to say that for vast classes of galaxy samples,
this factor of $(5s_f - 2)/b_f$ could be easily of order unity.
We will discuss this point in detail in \S\ref{Dependence on the bias and number count slope}.

In all illustrative examples below, we will consider for simplicity a gaussian selection function of width $\sigma$ 
\be
W(z,z_{0})=\frac{1}{\sqrt{2\pi}\sigma}\exp\left[-\frac{(z-z_{0})^2}{2\sigma^2}\right]\ ,
\ee
and the following cosmological parameters: the Hubble constant $h=0.7$, matter density $\Omega_m =0.27$, cosmological constant $\Omega_\Lambda =0.73$, baryon density $\Omega_b =0.046$, power spectrum slope $n=0.95$ and normalization $\sigma_8=0.8$. We employ the transfer function of \cite{Eisenstein_Hu:1997} and the prescription of \cite{Smith:2002} for the nonlinear power spectrum.

\section{Correction due to magnification bias}\label{Correction due to magnification bias}
\begin{figure}[h!]
\begin{center}
\includegraphics[width=0.85\linewidth]{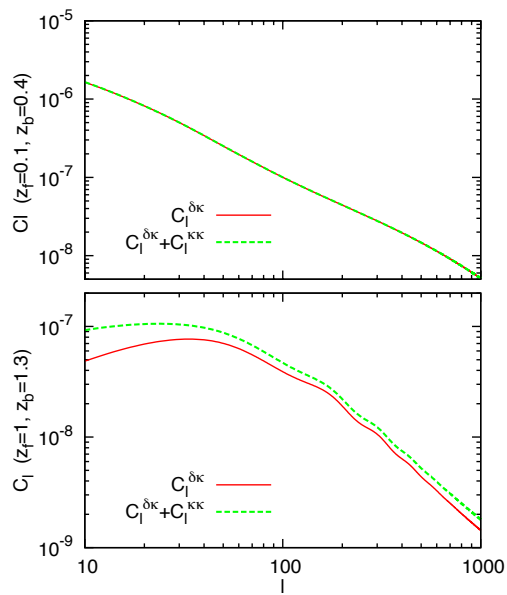}
\caption{Galaxy-lensing power spectrum $C^{\delta\kappa}_{\ell}(z_{f},z_{b})$ and the corrected power spectrum $C^{\delta\kappa}_{\ell}(z_{f},z_{b})+C^{\kappa\kappa}_{\ell}(z_{f},z_{b})$ for two redshift configurations. The redshift width of the selection function is $\sigma =0.07$.}
\label{Pgmu and Pmumu}
\end{center}
\end{figure}
We would like to understand better to what extent the correction due to magnification bias should be taken into account, and if so, in what cases. Before going into details, we would like to emphasize that the correction to $C^{\delta\kappa}_{\ell}$ due to $C^{\kappa\kappa}_{\ell}$ can be \textit{non-negligible}; in Fig.~\ref{Pgmu and Pmumu} we plotted the correlation ``signal'' $C^{\delta\kappa}_{\ell}(z_{f},z_{b})$ along with the corrected observable $C^{\delta\kappa}_{\ell}(z_{f},z_{b})+C^{\kappa\kappa}_{\ell}(z_{f},z_{b})$ for two foreground and background redshifts. We see that the correction is very small for low foreground redshifts, but this is not the case for a higher $z_{f}$. 

In the following, we will study in detail the ratio in Eq.(\ref{def reff}), or its cousin $r$
in Eq.(\ref{def r}), and their dependence
on the different parameters.

\subsection{Dependence on the background galaxies redshift $z_{b}$}\label{Dependence on the background galaxies redshift zb}
\begin{figure}[htbp]
\begin{center}
\includegraphics[width=0.85\linewidth]{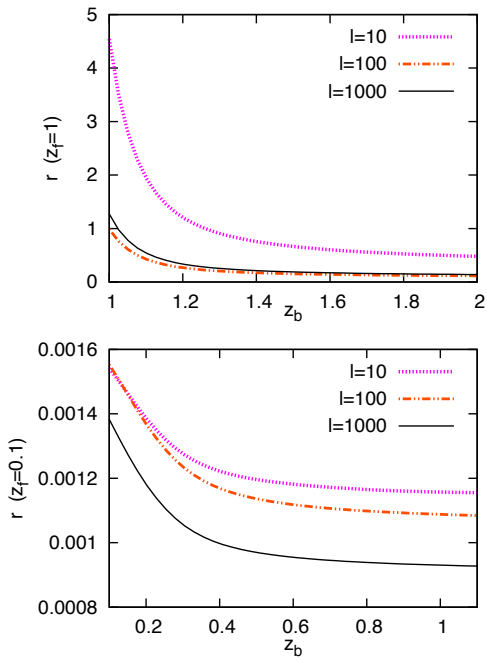}
\caption{ $r(z_{f},z_{b};\ell)=C^{\kappa\kappa}_{\ell}(z_{f},z_{b})/C^{\delta\kappa}_{\ell}(z_{f},z_{b})$ for two different fixed foreground redshifts $z_{f}=0.1,1$ and for three values of $\ell=10,100,1000$. The width of the selection function is $\sigma =0.07$. For each $z_{f}$ we let $z_{b}$ vary in $ [z_{f},z_{f}+1]$.
We can see that the ratio $r$ decreases with the separation between the background and foreground samples.}
\label{Ratio_r_funct_zb}
\end{center}
\end{figure}

The ratio $r$ depends on $z_{b}$ through the lensing weight function $g$. In Fig.~\ref{Ratio_r_funct_zb}, we 
can see that $r$ grows when the background sample gets closer to the foreground galaxies. 
This is easy to understand: the galaxy-lensing correlation, which is proportional to $C^{\delta\kappa}_\ell$,
becomes small when the source-lens separation becomes small.
More quantitatively, if one approximates the selection function by a delta function i.e. $W(z,z_f)\approx \delta(z-z_f)$ (which is 
a reasonnable approximation if the width $\sigma$ of the selection functions is small and if the sources and lenses distributions do not overlap), we have:
\be\label{C g kappa avec delta}
\begin{aligned}
C^{\delta\kappa}_{\ell}(z_{f},z_{b})\approx&\left(\frac{3}{2}\frac{H_{0}^{2}}{c^{2}}\Omega_{m,0}\right)\frac{\chi(z_b)-\chi(z_f)}{\chi(z_f)\chi(z_b)}\\
 &\qquad\qquad
\times(1+z_f)P(\ell/\chi(z_f),z_f)\ ,
\end{aligned}
\ee
showing that $C^{\delta\kappa}_{\ell}$ goes to zero for close background and foreground samples; this is due to the fact that when the lenses are very close to the sources, the lensing weight function $g$ is almost null (it would
exactly vanish if indeed $W$ were a perfect delta function).
Under the same approximation, we have
\be\label{C kappa kappa avec delta}
\begin{aligned}
C^{\kappa\kappa}_{\ell}(z_{f},z_{b}) &\approx \left(\frac{3}{2}\frac{H_{0}^{2}}{c^{2}}\Omega_{m,0}\right)^{2}\int_{0}^{z_{f}}\!\!dz  \frac{c}{H(z)}(1+z)^{2}\\
&\times\frac{\chi(z_{f})-\chi(z)}{\chi(z_{f})}\frac{\chi(z_{b})-\chi(z)}{\chi(z_{b})} P(\ell/\chi(z),z)
\end{aligned}
\ee
which tends toward a finite value when $z_{b}\rightarrow z_{f}$. It is therefore not surprising
to see that the ratio $r$ becomes large when the background galaxy redshift approaches that of the foreground. 
Of course, for realistic observations, the selection function is never quite a delta function, and the precise behavior at low background redshift will depend on its precise shape. We will study that aspect in more detail in \S \ref{Dependence on the multipole l} (see also \cite{LoVerde_Corrections:2007}).
Note also that when $z_b$ approaches $z_f$, the whole premise of galaxy-magnification cross-correlation
measurements breaks down: as described in \S \ref{gmcorr}, the idea of such measurements is to
correlate the number counts of a foreground sample and a background sample which are sufficiently
far apart so that the intrinsic clustering term $w_{gg}$ is unimportant.

Another interesting feature of Fig.~\ref{Ratio_r_funct_zb} is that for high $z_b$, $r$ asymptotes to some minimal value (independent of the background position). This can also be understood easily through the equations (\ref{C g kappa avec delta}),(\ref{C kappa kappa avec delta}); indeed, when $\chi(z_{f}) \ll \chi(z_b)$, $\chi(z_{b})$ cancels out of Eq.(\ref{C g kappa avec delta}) while Eq.(\ref{C kappa kappa avec delta}) becomes weakly dependent
on $\chi(z_b)$. Therefore, if one's goal is to minimize  (magnification bias induced) corrections to 
the galaxy-lensing correlations, it is important to make sure the sources and the lenses are widely
separated. However, it is worth noting that according to Fig.~\ref{Ratio_r_funct_zb}, 
$r$ can still be fairly substantial
even when $z_b$ differs significantly from $z_f$. This holds
especially if $z_f$ is sufficiently large. 
We will explore this further next.

\subsection{Dependence on the foreground galaxies redshift $z_{f}$}
\begin{figure}[htb]
\begin{center}
\includegraphics[width=0.9\linewidth]{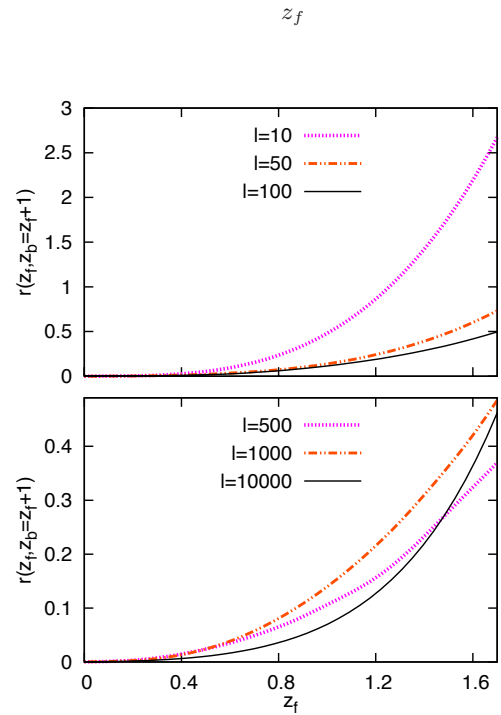}
\caption{Ratio $ r(z_{f},z_{b};\ell)$ for different foreground redshifts $z_{f}$; for each $z_{f}$, we choose a background redshift $z_{b}=z_{f}+1$, such that according to Fig.~\ref{Ratio_r_funct_zb}, $r$ is minimized.  The 
redshift width of the selection function is $\sigma =0.07$. On the top panel, low multipoles are shown, while the bottom plot shows $r$ for high multipoles.}
\label{Ratio_r_funct_zf}
\end{center}
\end{figure}
The foreground redshift appears to be the crucial parameter which determines 
when the correction to $C^{\delta\kappa}_{\ell}$ due to $C^{\kappa\kappa}_{\ell}$ is large. 
Indeed, the correlation $C^{\kappa\kappa}_{\ell}$ comes from the lensing of both background and foreground galaxies by the mass located between $z_{f}$ and $z=0$. The lensing efficiency associated with this mass
increases as $z_f$ increases.
This effect is already visible in Fig.~\ref{Ratio_r_funct_zb}, but is even more striking in Fig.~\ref{Ratio_r_funct_zf} in which we plot $r$ as a function of $z_{f}$ (for each point, we choose $z_{b}=z_{f}+1$ so that
$r$ is close to its asymptotic minimum value for this $z_f$).
Obviously, the curves for low multipoles have a very different behavior from the curves for high multipoles (we 
will study in more detail this dependence on $\ell$ below). However, the main point is that for each multipole $\ell$, $r$ grows with $z_{f}$ and reaches a non-negligible value for high $z_{f}$. For example, for  $\ell=1000$, $r=C^{\kappa\kappa}_{\ell}/C^{\delta\kappa}_{\ell}$ is above $1\%$ as soon as $z_{f}\gtrsim 0.37$, and above $5\%$ for $z_{f}\gtrsim 0.67$. 
As a consequence, magnification bias correction should be taken into account when interpreting
galaxy-lensing correlation measurements if the foreground redshift is sufficiently high.

This dependence on the foreground redshift $z_{f}$ also allows us to infer that this correction is negligible for the cross-correlations that have been observed until now. Indeed, in the case of the recently observed galaxy-shear correlations (or so called galaxy-galaxy lensing) using the SDSS survey \cite{Sheldon:2004,Sheldon:2007_I,Johnston:2007_II,Sheldon:2007_III,Mandelbaum:2005,Mandelbaum:2006}, the lenses (foreground)  are mostly distributed \cite{Sheldon:2004} around a mean redshift $\left<z_{f}\right>=0.1$. It is safe to assume that
these galaxies have $|5s_f - 2|/b_f$ of at best unity \cite{Sheldon:2004}. Therefore
the fractional correction is small: $|C^{\mu\kappa}_{\ell}/ C^{g\kappa}_{\ell}| \lesssim 1\%$.
Similar conclusions hold in the case of galaxy-magnification correlations measured by cross-correlating
foreground galaxy counts with background quasar counts, as carried out by \cite{Scranton_SDSS:2005}: here again, the distribution of the foreground galaxies is peaked around the low redshift $\left<z_{f}\right>=0.24$, leading to a less than $5\%$ correction even for a large number count slope.

The situation will be quite different for future weak lensing surveys, like LSST \cite{LSST:2006} or DUNE (now EUCLID) \cite{Dune_spie}; indeed, these surveys will obtain precise photometric redshifts up to very high redshifts $z\sim3$, and therefore \textit{the magnification bias correction will not be a priori negligible}. In particular, for high lens redshift $z_{f}\gtrsim1$, if the foreground number slope $s_{f}$ is not such that the magnification is small, the correction can be of \emph{order one}.
To keep the magnification bias correction small, one can either focus on low $z_f$ galaxies, or
select samples where $s_f$ is close to $0.4$ (or weigh galaxies by $1/(5s_f - 2)$). 
Perhaps a better strategy is to include the magnification
bias correction from the beginning when interpreting measurements.

\subsection{Dependence on the multipole $\ell$ and the selection function width $\sigma$}\label{Dependence on the multipole l}
\begin{figure}[!h]
\begin{center}
\includegraphics[width=0.9\linewidth]{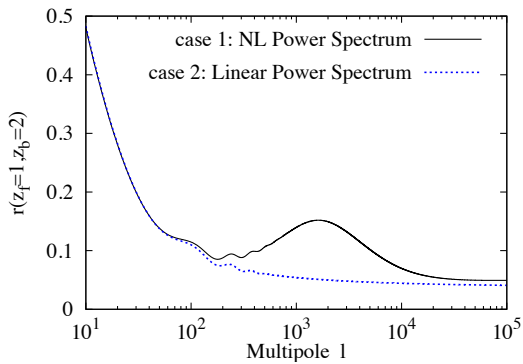}
\caption{Ratio $r(z_{f},z_{b};\ell)=C^{\kappa\kappa}_{\ell}(z_{f},z_{b})/C^{\delta\kappa}_{\ell}(z_{f},z_{b})$ as a function of the multipole $\ell$ for $z_{f}=1,z_{b}=2$ in two cases. \textit{Case 1}: $r$ is computed using a non-linear matter power spectum \cite{Smith:2002};  \textit{Case 2}:  $r$ is computed assuming the matter power spectrum is equal to the linear one. For low $\ell$, $r$ decreases steeply, before reaching a roughly constant plateau for high $\ell$. }
\label{Ratio_r_funct_l}
\end{center}
\end{figure}

As discussed above, the correction due to $C^{\kappa\kappa}_{\ell}$ depends on $\ell$; in Fig.~\ref{Ratio_r_funct_l}, we plot $r$ as a function of $\ell$ for $z_{f}=1,z_{b}=2$ in the case of non-linear matter power spectum (case 1); in order to emphasize the effect of non-linearities, we also plot $r(\ell)$ assuming the matter power spectrum is equal to the linear one (case 2). (All figures in the rest of this paper use the nonlinear power
spectrum.)

We can see that $r$ generally decreases with $\ell$. We can also see a bump around $\ell\approx 1000$ in the non-linear case. Since non-linearities become important at different $\ell$'s for $C_\ell^{\kappa\kappa}$
and $C_\ell^{\delta\kappa}$ (because the former is sensitive to fluctuations between redshift
zero and $z_f$ while the latter is sensitive to fluctuations at around $z_f$), it is not surprising
that their ratio is not monotonic.
At sufficiently high $\ell$'s, the ratio $r$ asymptotes to some low value.

The overall behavior of $r$ can be understood as follows. Approximating once again the
selection function as a delta function, we need only to consider the ratio of 
Eq.(\ref{C kappa kappa avec delta}) to Eq.(\ref{C g kappa avec delta}). 
At sufficiently high $\ell$'s, the integral over $z$ in Eq.(\ref{C kappa kappa avec delta})
is such that it is dominated by $z$'s where $P(\ell/\chi(z), z)$ is more or less power-law
in $\ell$ i.e. $P \propto \ell^{-3}$ (originating from the fact that
$P(k) \propto k^{-3}$ at high $k$'s). Eq.(\ref{C g kappa avec delta}) is likewise
roughly proportional to $\ell^{-3}$ for sufficiently high $\ell$'s.
Their ratio $r$ is therefore independent of $\ell$, giving rise to the high $\ell$ asymptote
that we see in Fig.~\ref{Ratio_r_funct_l}.
At low $\ell$'s, on the other hand, the integral in Eq.(\ref{C kappa kappa avec delta})
goes through the maximum of the matter power spectrum, and the result depends strongly on $\ell$, 
leading to the steep decreasing slope for low $\ell$ noticed in Fig.~\ref{Ratio_r_funct_l}.

We should emphasize that at high $\ell$'s ($\gtrsim 10^3$), the actual ratios of interest
$C^{g\kappa}_\ell/C^{\mu\kappa}_\ell$ and $C^{g\mu}_\ell/C^{\mu\mu}_\ell$ 
(Eqs. [\ref{Cggamma}], [\ref{Cmugamma}], [\ref{Cgmu}], [\ref{Cmumu}])
are affected by nonlinear and scale dependent galaxy bias.


\begin{figure}[!h]
\begin{center}
\includegraphics[width=0.9\linewidth]{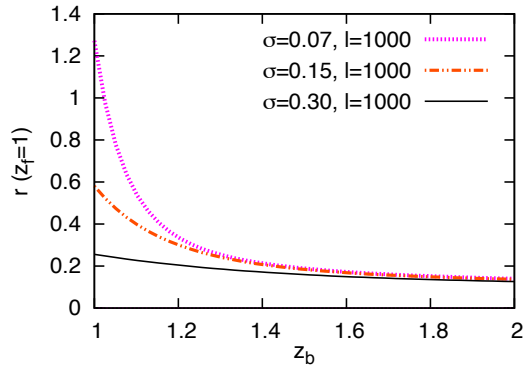}
\caption{$r(z_{f},z_{b};\ell)=C^{\kappa\kappa}_{\ell}(z_{f},z_{b})/C^{\delta\kappa}_{\ell}(z_{f},z_{b})$ as a function of $z_{b}$ for $z_{f}=1$ and $\ell=1000$; three different widths of the selection function are shown: $\sigma =0.07,0.15,0.30$. For background redshifts close to foreground ones, $r$ depends strongly on $\sigma$, decreasing as $\sigma$ increases; for high $z_{b}\gtrsim z_{f}+3\sigma$, the width of selection function has no influence anymore, as expected from the delta function approximation.}
\label{Ratio_r_funct_zb_sigma}
\end{center}
\end{figure}

In this paper, we are mostly interested in the cases where the foreground and background distributions do not overlap, and for which the width $\sigma$ of the selection function is of little importance, as long as it is small (the case where the foreground and background galaxies are at the same redshift has been studied extensively in \cite{LoVerde_Corrections:2007}). This is confirmed by Fig.~\ref{Ratio_r_funct_zb_sigma}, where we see
that as long as the separation between foreground and background is large compared to $\sigma$,
the precise value of $\sigma$ does not matter much.
In cases where $z_b$ is actually close to $z_f$, the main effect of increasing $\sigma$ is to
make $C^{\delta\kappa}_\ell$ larger, hence making the ratio $r$ smaller (recall that
$C^{\delta\kappa}_\ell$ vanishes if the selection function is an exact delta function).

\subsection{Dependence on the bias $b_{f}$ and number count slope $s_{f}$}\label{Dependence on the bias and number count slope}
In the previous sections
, we have studied the parameter dependence of the ratio $
r={C^{\kappa\kappa}_{\ell}}/{C^{\delta\kappa}_{\ell}}$.
However, the quantity of observational interest is actually
$\frac{(5s_{f}-2)}{b_{f}}\; r$ (Eq.(\ref{def reff})), which depends crucially on the lens sample dependent quantities ${(5s_{f}-2)}$ and ${b_{f}}$. The ratio $(5 s_f - 2)/b_f$ can vary anywhere from $-1$ to $2$
depending on the sample limiting magnitude and redshift, and it generally increases as 
one focuses on brighter galaxies and higher redshifts (see 
Fig.~1 of \cite{Hui_Anis_I:2007}). 
It is worth noting that measurements that require spectroscopic redshifts (such as in the geometrical method
proposed by \cite{Jain_T:2004}, discussed below) focus on bright galaxies for which
$(5 s_f - 2)/b_f$ is generally large i.e. order unity or above.

\section{Geometrical methods and magnification bias}\label{Geometrical methods}
\begin{figure}[!h]
\begin{center}
\includegraphics[width=0.85\linewidth]{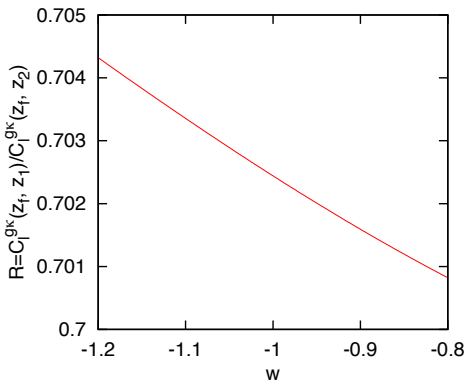}
\caption{Dependence of the ratio $R=C^{g\kappa}_{\ell}(z_{f},z_{1})/C^{g\kappa}_{\ell}(z_{f},z_{2})$ on the DE equation of state $w$ for $z_{f}=0.3,z_{1}=0.5,z_{2}=0.7$. Note that $R$ is a weak function of $w$.}
\label{Ratio_JT}
\end{center}
\end{figure}

By "geometrical methods", we refer to techniques that involve combining lensing measurements at different
redshifts to obtain measures that are purely geometrical in nature (i.e. they depend only
on combinations of geometrical distances, and not on the power spectrum or its growth).
There are two simple observations we would like to make about these techniques.
First, there is a simple way to implement these techniques using galaxy counts alone,
without using any galaxy shape information. Second, magnification bias corrections may affect these measures
in a significant way, or more precisely, affect significantly the inferrence of dark energy parameters from these measures.

\subsection{Extensions using number counts}
\label{extension}

Jain and Taylor \cite{Jain_T:2004} (see also \cite{Bernstein:2003,Taylor:2006})
proposed using the ratio of galaxy-shear correlations (i.e. in a galaxy-galaxy
lensing experiment, as described in \S \ref{gshearcorr})
to constrain a certain combination of angular diameter distances:
\be\label{ratio R}
R=\frac{C^{g\kappa}_{\ell}(z_{f},z_{1})}{C^{g\kappa}_{\ell}(z_{f},z_{2})}=\frac{\left(\chi(z_1)-\chi(z_f)\right)\chi(z_2)}{\left(\chi(z_2)-\chi(z_f)\right)\chi(z_1)}\ .
\ee
The second equality holds if the selection function are sharply peaked; this places
an especially high demand on the redshift accuracy of the foreground galaxies \cite{Bernstein:2003}.

It is fairly straightforward to see that the same ratio can be obtained by
galaxy-magnification correlation measurements, the type described in \S \ref{gmcorr}.
In other words, by cross-correlating the number counts of a foreground sample
and a background sample of galaxies, one can isolate $C^{g\mu}_\ell$ (Eq.[\ref{Cgmu}], temporarily ignoring
the magnification bias correction that is in Eq.[\ref{Cmumu}]). If one can measure this cross-correlation using
two different background samples but with the same foreground, we can 
form a ratio analgous to the Jain and Taylor ratio of Eq.(\ref{ratio R}):
\be\label{ratio R new}
R_{\rm new}=\frac{C^{g\mu}_{\ell}(z_{f},z_{1})}{C^{g\mu}_{\ell}(z_{f},z_{2})}
=\frac{(5s_{1}-2)\left(\chi(z_1)-\chi(z_f)\right)\chi(z_2)}{(5s_{2}-2)\left(\chi(z_2)-\chi(z_f)\right)\chi(z_1)}\
\ee
where the second equality holds again under the assumption of a sufficiently narrow
selection function. This ratio differs from Eq.(\ref{ratio R}) only by factors that involves the
number count slope which can be independently measured.

Similar logic applies to generalizations of the Jain and Taylor ratio, such as
the one proposed by Zhang et al. \cite{Zhang_Hui:2003}. It exploits what is
known as an offset linear scaling of the galaxy-shear correlation (similar scaling
applies to shear-shear correlation as well):
\be
C^{g\kappa}_{\ell}(z_{f},z_{b}) =F_{\ell}(z_{f})+{G_{\ell}(z_{f})}/{\chi_{\text{eff}}(z_{b})}
\ee
where
\be
\frac{1}{\chi_{\text{eff}}(z_{b})}=\int_{0}^{\infty}dz'\;\frac{W(z',z_{b})}{\chi(z')}
\ee
and
\bea
F_{\ell}(z_{f}) &=& \left(\frac{3}{2}\frac{H_{0}^{2}}{c^{2}}\Omega_{m,0}\right)b_{f}\nonumber\\
&\times&\int_{0}^{\infty}dz\; \frac{W(z,z_{f})}{\chi(z)}(1+z)P(\ell/\chi(z),z)\quad\quad\\
G_{\ell}(z_{f})& =& -\left(\frac{3}{2}\frac{H_{0}^{2}}{c^{2}}\Omega_{m,0}\right)b_{f}\nonumber\\
&\times&\int_{0}^{\infty}dz\; W(z,z_{f})(1+z)P(\ell/\chi(z),z).
\eea
Zhang et al. \cite{Zhang_Hui:2003} suggested combinations of $C^{g\kappa}_\ell$'s from different samples
which are purely geometrical in nature, such as
\be
R^Z\equiv\frac{C^{g\kappa}_{\ell}(z_{f},z_{3})-C^{g\kappa}_{\ell}(z_{f},z_{1})}{C^{g\kappa}_{\ell}(z_{f},z_{2})-C^{g\kappa}_{\ell}(z_{f},z_{1})}\!=\!\frac{\chi_{\text{eff}}(z_{3})^{-1}\!-\!\chi_{\text{eff}}(z_{1})^{-1}}{\chi_{\text{eff}}(z_{2})^{-1}\!-\!\chi_{\text{eff}}(z_{1})^{-1}}
\ee
An advantage of this approach compared to \cite{Jain_T:2004} is that it is less demanding
on the redshift accuracy.

Once again, one can implement the same idea using the cross-correlation of number counts alone
i.e. one can study instead:
\bea
R^Z_{\rm new}&\equiv&\frac{C^{g\mu}_{\ell}(z_{f},z_{3})/(5s_3 - 2) - C^{g\mu}_{\ell}(z_{f},z_{1})/(5s_1 - 2)}{C^{g\mu}_{\ell}(z_{f},z_{2})/(5s_2 - 2)-C^{g\mu}_{\ell}(z_{f},z_{1})/(5s_1 - 2)}\nonumber\\
&=&\frac{\chi_{\text{eff}}(z_{3})^{-1}-\chi_{\text{eff}}(z_{1})^{-1}}{\chi_{\text{eff}}(z_{2})^{-1}-\chi_{\text{eff}}(z_{1})^{-1}}
\eea

\subsection{Magnification bias corrections}
\label{magbiascorr}
\begin{figure}[!h]
\begin{center}
\includegraphics[width=0.9\linewidth]{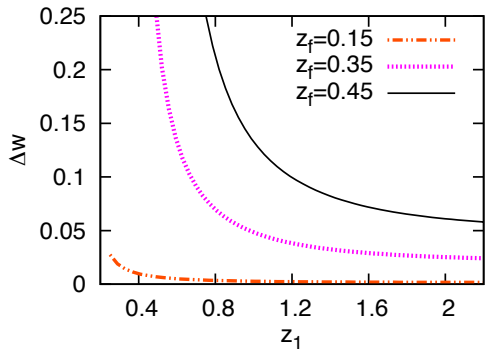}
\caption{Bias in the inferred dark energy equation of state $\Delta w$ if magnification
bias corrections are ignored in interpreting the Jain and Taylor ratio. This is
as a function of the background redshift $z_1$, for 
three foreground redshifts $z_{f}=0.15,0.35,0.45$. For each $z_{1}$, we fix the other
background redshift $z_{2}=z_{1}+0.4$ (our results do not
depend sensitively on the choice of $z_2 - z_1$).
The selection functions are approximated by delta functions, and $(5s_f - 2)/b_f \sim 1$ is assumed. The calculation is done for $\ell = 1000$.}
\label{JT_zb_dependence}
\end{center}
\end{figure}
The methods described above are interesting because they allow us
to isolate from cross-correlation measurements purely geometrical constraints on dark energy. 
Such constraints can be compared with constraints from the growth of structure as a consistency
test of general relativity. Fig.~\ref{Ratio_JT} shows the Jain and Taylor ratio $R$ as a function
of the dark energy equation of state $w$. We would like to draw attention to the fact
that $R$ is a weak function of $w$. This means an accurate determination of $w$ would require
a very precise measurement of $R$. For instance, in the configuration of Fig.~\ref{Ratio_JT}, to constrain $w$ to $5 \%$, we would need
to measure $R$ to within $\sim 0.04 \%$. This suggests the magnification bias
corrections we have been studying might be especially relevant: their presence, if unaccounted for,
would lead to incorrect inferrence on the dark energy equation of state.

More quantitatively, the {\it observed} Jain and Taylor
ratio is
\bea
R_{\rm obs} &=&\frac{C^{g\kappa}_{\ell}(z_{f},z_{1})+C^{\mu\kappa}_{\ell}(z_{f},z_{1})}{C^{g\kappa}_{\ell}(z_{f},z_{2})+C^{\mu\kappa}_{\ell}(z_{f},z_{2})}\nonumber\quad\\
&=&\frac{C^{g\kappa}_{\ell}(z_{f},z_{1})}{C^{g\kappa}_{\ell}(z_{f},z_{2})}\left(\frac{1+\frac{(5s_{f}-2)}{b_{f}}\; r(z_{f},z_{1};\ell)}{1+\frac{(5s_{f}-2)}{b_{f}}\; r(z_{f},z_{2};\ell)}\right)\qquad\label{ratio R corrected}
\eea
where $z_f$ is the foreground redshift, and $z_1$ and $z_2$ are two different background redshifts.
Inferring $w$ from $R_{\rm obs}$, without accounting for its difference from the
idealized ratio $R$ in Eq.(\ref{ratio R corrected}) (i.e. using $R$ as the model for the data),
leads then to an inferred DE equation of state $w_{\rm inferred}$ which differs from the true one  $w_{\rm true}$; we call $\Delta w=w_{\rm true}-w_{\rm inferred}$ the resulting bias.
In Fig.~\ref{JT_zb_dependence}, this bias $\Delta w$ is shown for various redshift configurations and $\ell=1000$. Here, we consider redshifts ranging up to $z = 2.6$, approximate the selection functions by delta functions, and assume $(5s_f - 2)/b_f \sim 1$. The approximation and assumption
are reasonable given that
the Jain and Taylor method requires essentially spectroscopic redshifts for
the foreground galaxies: such galaxies are generally bright and their values for $(5s_f - 2)/b_f$
are typically large (see \cite{Hui_Anis_I:2007}).
We find that for very low foreground redshifts ($z_{f}\lesssim0.15$), the bias $\Delta w$
is generally less than $5 \%$ for all combinations of the two background redshifts $z_1$ and $z_2$.
For intermediate redshifts $0.15\lesssim z_{f}\lesssim 0.45$, whether one can 
keep $\Delta w$ below $5\% $ depends on the $z_{1},z_{2}$ configuration: 
for low $z_{1},z_{2}$ (i.e. approaching $z_f$), 
the magnification bias correction is simply too large to be ignored;
for sufficiently high $z_1, z_2$, $r(z{f}, z_1; \ell)$ and $r(z{f}, z_2; \ell)$ becomes
close to each other, {\it almost} independent of the background redshifts
(see Fig.~\ref{Ratio_r_funct_zb}) and so the magnification bias corrections cancel each other
to a large extent in $R_{\rm obs}$. 
Finally, for high foreground redshifts ($z_{f}\gtrsim0.45$), the magnification bias 
corrections are simply too large to allow such delicate cancelations: 
$\Delta w$ exceeds $5 \%$ for all background configurations.

One could of course, when interpreting data, 
account for the magnification bias corrections from the beginning
and avoid such a bias $\Delta w$. Note however that these corrections depend on the power spectrum,
making the observed ratio $R_{\rm obs}$ not strictly geometrical anymore.
In other words, this ratio is a useful, pure geometrical measure only in so far as
the magnification bias corrections are small.

\section{Conclusion}\label{Conclusion}

In this paper, we present a detailed study of the correction due to the magnification bias of foreground galaxies in the context of galaxy-lensing cross-correlations. We find that the most important factor that determines the size of this correction
is the foreground redshift: the size increases with redshift.
For example, we found that for $\ell=1000$, the correction is above $1\%\times (5s_{f}-2)/b_{f}$ for $z_{f}\gtrsim 0.37$, and above $5\%\times (5s_{f}-2)/b_{f}$ for $z_{f}\gtrsim 0.67$. 
Maintaining a low foreground redshift is therefore necessary to keep the magnification bias correction small.
(We find a sufficently large separation between foreground and background, $\Delta z \sim 1$, is also necessary.)
Or else, one should explicitly account for the magnification bias correction when interpreting data. This is
especially true for future surveys which are expected to reach high precision and high redshifts.
Finally, we present a simple extension of existing geometrical methods
to measure dark energy parameters independently of the matter power spectrum.
In the case of galaxy-shear \cite{Jain_T:2004,Zhang_Hui:2003} and shear-shear \cite{Zhang_Hui:2003} correlation functions: the galaxy-shear correlation can be replaced by galaxy-magnification correlation, allowing measurements that involve only galaxy counts, with no galaxy shape
information necessary. However, we also point out that
these geometrical methods are generically sensitive to magnification bias corrections.
For $(5s_{f}-2)/b_{f}=1$, we find that the foreground redshift should be limited to
$z_f \lesssim 0.45$ in order to avoid biasing the inferred dark energy equation of state
$w$ by more than $5 \%$, and that even for a low foreground redshift ($< 0.45$),
the background samples must be well separated from the foreground to avoid incurring a bias of similar magnitude.

As this paper was nearing completion, we became aware of a preprint by
Bernstein \cite{Bernstein:2008aq} that has some overlap with the magnification bias issues
that we discuss here.

\section*{Acknowledgments}

Research for this work is supported in part by the DOE, grant DE-FG02-92-ER40699,
and the Initiatives in Science and Engineering Program
at Columbia University.


\bibliographystyle{apsrev}
\bibliography{bibliography}

\end{document}